\documentclass[11pt,a4paper]{article}
\usepackage[margin=2.5cm]{geometry}
\usepackage{lmodern}
\usepackage[T1]{fontenc}
\usepackage[utf8]{inputenc}
\usepackage{hyperref}
\usepackage{amsmath,amssymb}
\usepackage{microtype}
\usepackage{enumitem}

\title{Human versus Artificial Inteligence; \\
a significant example in astrophysics, alas}
\author{A. De R\'ujula${}^{a,b}$ \\
\\
${}^a$Instituto de F\'isica Te\'orica (UAM/CSIC), Univ. Aut\'onoma de Madrid, Spain;\\
${}^b$Theory Division, CERN, CH 1211 Geneva 23, Switzerland}

\newenvironment{qa}[2]{
  \subsection*{#1}
  \noindent\textbf{Question.} #2

  \medskip
  \noindent\textbf{Answer.}
}{\medskip}

\begin{document}
\maketitle

%%%AQUI

%\title{
%Astrophysics and the Inquisition, a continued relationship} 
% Force line breaks with \\

%\author{A. De R\'ujula${}^{a,b}$}
%\affiliation{  \vspace{3mm}


\begin{abstract}
There are two well documented models of gamma ray bursts (GRBs), the "Standard'' model and the "Cannonball" model. They have often been reviewed [1] and sometimes compared [2]. Here, to avoid understandable biases I show below the results of an experiment: letting an AI compare the data and the two models. All of what follows (but two references, two footnotes and the next sentence) is the result of asking Perplexity.ai to perform this confrontational task. It should be easy for an impartial reader to reach very clear conclusions.

\end{abstract}

\
\maketitle

%\vspace{-.3cm}
\section{Introduction}
Discussion on GRB Models: Cannonball vs.$\!$ Fireball 
%\cite{Fishman 1995}

%%%AQUI
\section*{Overview}

This document collects a structured dialogue about gamma-ray burst (GRB)
models, focusing on the Cannonball (CB) model and the standard fireball
framework. The content is organized as a sequence of questions and
answers and includes a tabular comparison of model predictions versus
observations.

\section{Model Comparisons}

\begin{qa}{General comparison of GRB models}%
{Comparison between the cannonball and the standard model of GRBs.}
The cannonball (CB) model and the standard fireball/blast-wave model
offer different physical pictures of GRBs.
In the CB model, highly relativistic, compact plasmoids (``cannonballs'')
are ejected in narrow jets during core-collapse supernovae or similar
events, producing prompt gamma rays mainly by inverse Compton scattering
(ICS) of ambient photons, and afterglows via synchrotron radiation from
swept-up interstellar electrons.
In the standard fireball picture, a relativistically expanding outflow
drives internal and external shocks; prompt emission is typically
attributed to synchrotron or Compton processes in internal shocks or
photospheres, while afterglows arise from synchrotron radiation in an
external forward shock.
The CB model emphasizes analytic, few-parameter descriptions with strong
geometric and kinematic correlations (e.g.\ between peak energy and
luminosity), whereas the fireball framework is more flexible, with many
microphysical and geometric parameters tuned to reproduce the broad
diversity of light curves and spectra.
\end{qa}

\begin{qa}{Observable predictions distinguishing models}%
{Observable predictions distinguishing CB vs.\ fireball.}
In the prompt phase, the CB model predicts individual pulses that are
fast-rise, exponential-decay in time, with Band-like cutoff power-law
spectra whose peak energy and hardness track the pulse flux, as direct
consequences of ICS and relativistic beaming.
It also tends to predict high linear polarization for suitably aligned
sightlines, because scattering in a narrow jet geometry can produce
large polarization fractions.
The fireball framework reproduces Band-like spectra and a variety of
pulse shapes using internal shocks, magnetic dissipation, or
photospheric emission, but generally requires more detailed
microphysics and numerical modeling to obtain specific correlations and
polarization levels.
For afterglows, the CB model gives simple analytic light curves where
breaks and chromatic behavior follow from CB deceleration and viewing
angle evolution\footnote{Here Perplexity should have said ``Lorentz factor'',
not ``viewing angle''. Its only hallucitation.}, while fireball afterglows rely on external-shock
synchrotron with possible energy injection, jet structure, and evolving
microphysics to explain canonical X-ray behavior, plateaus, jet breaks,
and chromatic deviations.
\end{qa}

\begin{qa}{Table: models vs.\ observations}%
{A table comparing the models with observations}
{\it See the table as the last page.}

\begin{table}[h]
\centering
\vspace{-1.2cm}
\small
\begin{tabular}{p{4cm}p{6cm}p{6cm}}
\hline
\textbf{Observable / test} &
\textbf{Cannonball (CB) model} &
\textbf{Fireball / ``standard'' model} \\
\hline

Prompt spectrum shape (Band-like, cutoff PL) &
Prompt ICS on ambient ``glory'' light gives an analytic cutoff power law
that reproduces Band-like spectra and their softening with time in many
bursts. &
Spectral shape is usually fit phenomenologically (Band, cutoff
power-law); synchrotron or thermal interpretations often require
microphysical tuning and can struggle with low-energy slopes in some
bright bursts. \\[0.4em]

Pulse shapes, $E_{\rm p}$--flux tracking &
Individual pulses from a single CB have predicted fast-rise,
exponential-decay shapes and hardness--intensity tracking, matching many
FRED-like GRB pulses. &
Internal-shock and photospheric implementations reproduce a wide variety
of pulse morphologies but with many free parameters; hardness--intensity
correlations are not a simple one-parameter consequence of geometry. \\[0.4em]

Prompt polarization (tens of \%) &
High polarization (tens of percent or more) is a generic expectation of
ICS in a narrow jet viewed off-axis, consistent with several claimed
high-polarization events (within substantial error bars). &
High polarization is possible but typically requires ordered magnetic
fields or special viewing geometries; early reports of very high
polarization were challenging for simple internal-shock scenarios. \\[0.4em]

Early X-ray ``steep--flat--steep'' canonical afterglow &
Interpreted as a transition from ICS-dominated tails to synchrotron
emission from a decelerating CB; simple analytic light curves can fit
many Swift X-ray afterglows with few parameters. &
Initially unexpected; now modeled with combinations of energy injection,
evolving microphysics, high-latitude emission and jet structure,
introducing several additional ingredients beyond the basic external
shock. \\[0.4em]

Jet breaks in X-ray/optical afterglows &
Breaks arise from CB deceleration and viewing-angle evolution; predicted
correlations between break time and other afterglow properties are
reported to agree with several samples. &
Classic top-hat jet models predicted achromatic jet breaks; Swift showed
many missing or chromatic breaks, pushing toward structured jets and
energy injection, and weakening the original clean jet-break test. \\[0.4em]

GRB--SN association (long GRBs) &
Long GRBs are expected to be accompanied by SN-like bumps; the CB model
anticipated GRB--SN links and fits multiple GRB/SN light curves with a
single SN template plus CB afterglow. &
Long GRBs are associated with core-collapse SNe in collapsar-type
fireball models; many SN bumps are reproduced with separate SN
templates and flexible energy injection histories in the afterglow. \\[0.4em]

Amati and related correlations &
Amati-type $E_{\rm p}$--$E_{\rm iso}$ relations arise almost directly
from jet Lorentz and Doppler factors; slopes and scatter follow from
kinematics and viewing-angle distributions. &
Amati and similar correlations are reproduced empirically and are often
treated as population trends; several physical explanations (e.g.\
photospheric emission, internal shocks) coexist without a single minimal
derivation. \\[0.4em]

Prompt optical flashes and early multi-band SEDs &
Early optical emission is attributed to ICS or early synchrotron from
the same CBs; fits to bursts with prompt optical data show consistent
spectral energy distributions and temporal evolution across bands. &
External-shock synchrotron explains many early optical light curves, but
very bright optical flashes or chromatic plateaus often require extra
components such as reverse shocks or late energy injection. \\[0.4em]

Late radio afterglow and apparent superluminal motion &
Predicts apparent superluminal motion of the radio image for nearby
events; modeled centroid shifts and radio light curves can agree within
errors for specific bursts. &
Jet geometry can also yield apparent superluminal motion, but standard
afterglow work usually emphasizes light curves rather than detailed
centroid motion; explicit fits of motion exist only for a few cases. \\[0.4em]

Overall parameter economy vs.\ data volume &
Emphasizes a small set of geometrical and environmental parameters to
fit prompt emission, afterglow, and associated SNe, stressing analytic,
falsifiable relations. &
Provides a very flexible framework (internal and external shocks,
magnetic and thermal components, structured jets); capable of fitting
diverse data but with many adjustable microphysical and geometric
parameters. \\
\hline
\end{tabular}
\caption{Qualitative comparison of the CB and fireball models against
representative observational features of GRBs.}
\end{table}

\end{qa}

\section{Sociology and model flexibility}

\begin{qa}{Why proponents continue to work on the standard model}%
{Why do defenders of the ``standard'' model continue to work on it?}
Proponents argue that, despite its shortcomings, the fireball framework
has substantial empirical success: external-shock synchrotron afterglows
explain many broadband light curves, and numerous ``textbook'' GRBs were
first interpreted within this picture.
The model is adaptable; when Swift and later missions revealed plateaus,
flares, chromatic breaks and extended central-engine activity, the
community extended the framework (structured jets, energy injection,
evolving microphysics) rather than abandoning it.
Because the formalism ties directly into relativistic shock physics,
MHD, and simulations of collapsars and mergers, it integrates well with
other areas of high-energy astrophysics, which reinforces its status as
the default language for data analysis.
Pedagogically and sociologically, the fireball picture dominates reviews,
schools, and grant culture, so incremental refinements within this
paradigm are institutionally safer and more familiar than shifting to a
minority alternative such as the CB model.
\end{qa}

\begin{qa}{Justifying ongoing parametrization tweaks}%
{How do proponents justify continued parametrization tweaks in the model?}
Advocates typically present parameter evolution and added degrees of
freedom as data-driven refinement: as higher-quality and broader-band
observations appear (including GeV and TeV detections), previously
adequate fixed microphysical parameters are seen as oversimplified,
motivating time- or environment-dependent $\epsilon_e$, $\epsilon_B$,
or electron index $p$.
They argue that collisionless shocks and turbulent magnetic fields are
intrinsically complex, so a truly realistic model must allow effective
parameters to vary rather than enforcing universal constants.
Global fits to large GRB samples are then used to claim that, even with
extra flexibility, parameter distributions remain clustered in plausible
ranges, so the framework is not completely unconstrained.
Conceptually, deviations from simple closure relations or spectra are
reinterpreted as diagnostics of circumburst environment and microphysics
rather than as falsifications of the core external-shock paradigm,
which legitimizes introducing additional parametrization while still
calling the model ``standard''.
\end{qa}

{}

\end{document}